\documentclass[a4,amsmath,12pt]{article}

\usepackage{amssymb}
\usepackage{amsmath}
\usepackage{cite}
\usepackage{euscript}

\def\ZZ{\mathbb{Z}}
\def\RR{\mathbb{R}}
\def\r{\EuScript{R}}

\begin{document}

\titlepage
\begin{flushright}
hep-th/9903164 \\
LPTM 99/77\\
Saclay T99/028
\end{flushright}
\vskip 1cm
\begin{center}
{\bf \Large Super Fivebranes near the boundary of
$AdS_7\times S^4$}
\end{center}
\vskip 1cm
\begin{center}

C. Grojean$^{a}$\footnote{e-mail: grojean@spht.saclay.cea.fr} 
{\it and} J. Mourad$^{b,c}$\footnote{e-mail:
mourad@lyre.th.u-psud.fr}
\end{center} 
\vskip 0.5cm
\begin{center}
$^a$ {\it Service de Physique Th{\'e}orique, 
CEA-Saclay\\
F-91191 Gif/Yvette Cedex, France}\\
$^b$ {\it Laboratoire de Physique Th{\'e}orique et Mod{\'e}lisation,\\
Universit{\'e} de Cergy-Pontoise\\
Site Saint-Martin, F-95302 Cergy-Pontoise, France}\\
$^c$ {\it Laboratoire de Physique Th{\'e}orique 
\footnote{Unit{\'e} mixte de recherche --- UMR 8627.},\\
 Universit{\'e} de Paris-Sud, B{\^a}t. 211, F-91405 Orsay Cedex,
 France}
\end{center}
\vskip 2cm
\begin{center}
{\bf \large Abstract}
\end{center}

We  determine, to the first order in the radius 
of Anti-de-Sitter,
the realisation of the $OSp(6,2|2)$ superconformal algebra
on vector fields. We 
then calculate, to this order,
the superspace metric describing the background of 
$AdS_7\times S^4$. The coordinates we work with are adapted
to a 6+5 splitting of the eleven dimensional superspace.
Finally, we deduce in a manifestly supersymmetric
form the equations governing the dynamics of the
fivebrane near the boundary of $AdS_7$.

\noindent
\newpage

\section{Introduction}
\setcounter{equation}{0}

The relation between Anti-de-Sitter space (AdS)
and superconformal field theories defined on its boundary 
has been a subject of interest for a relatively long time
\cite{an,ber,gi}
( see \cite{duf} for a review).
The AdS/CFT conjecture \cite{mal, gub, wi} is the latest proposed
such relation and has, if correct, far reaching 
consequences: it allows, in principle, a non-perturbative
definition of string or M-theory on AdS backgrounds.  
The conjecture states \cite{mal}, for example, that
M-theory on $AdS_7\times S^4$ is equivalent to a
six-dimensional superconformal (2,0) field theory \cite{two}
defined on the boundary
of $AdS$. Little is known about M-theory
besides that its low energy approximation is given by
eleven-dimensional
supergravity, its compactification
on a circle leads to type IIA superstring \cite{wit2} and
that it contains membranes \cite{ber2} and 
fivebranes \cite{five,howe,aga,ban,fiveaction}.
The worldvolume action of super p-branes
and/or their equations of motion are formulated
in a superspace background. 
So the study of p-branes in
AdS space necessitates the determination of the 
superbackground. The bosonic $SO(6,2)\times SO(5)$
symmetry of $AdS_7\times S^4$ is promoted, when one considers 
the fermionic coordinates in addition to the bosonic ones,
to a $Osp(6,2|2)$ superconformal symmetry. 
The expression of the moving basis
and the rest of the background  depend on the 
eleven-dimensional 
supercoordinates one chooses to work with or equivalently
on the realisation of the isometries as super vector fields in
superspace. Such a realisation was found in 
\cite{DWPPS, DFFFTT, Claus, ck} using the coset
approach \cite{coset}. With a particular choice of super 
coordinates,
this approach leads to closed expressions for the isometries 
of $AdS_7\times S^4$ as well as the moving basis.
However these coordinates are not particularly adapted
to a 6+5 splitting of eleven dimensional spacetime, the six
dimensions representing the fivebrane's worldvolume
and the five remaining dimensions the transverse dimensions.
The aim of this paper is propose such coordinates and use 
the resulting isometries and background to
deduce in a manifestly supersymmetric 
form the equations of motion
of a fivebrane near the boundary of $AdS$.  

In order to study the super fivebrane in the $AdS_7\times S^4$
background there exists a privileged system of coordinates
which can be seen as follows: the fivebrane super
worldvolume is spanned by the (2,0) six-dimensional
supercoordinates ($x^\mu, \theta^{\alpha}$) 
where $\mu=0,\dots 5$ and $\alpha$ is a spin index of
the $(4_+,4)$ spinorial representation of $SO(5,1)\times
SO(5)$. The transverse fluctuations of the fivebrane
are described by a (2,0) multiplet which comprises five scalars
$\phi^i$ in the vectorial representation
of $SO(5)$ and a symplectic-Weyl-Majorana fermion 
$\psi^{\alpha'}$ in the $(4_-,4)$ spinorial representation of
$SO(5,1)\times SO(5)$. A convenient set of super-coordinates
to parametrise the eleven-dimensional superspace is thus given by
($x^\mu, \phi^i, \theta^{\alpha},
\psi^{\alpha'}$). The (2,0) super Poincar{\'e} six-dimensional 
group is a subgroup of $Osp(6,2|2)$ and so it has natural
realisation on the six-dimensional super worldvolume
of the fivebrane.  
In this paper, we shall use these coordinates to
parametrise the eleven-dimensional superspace.
This will allow us to find, in a manifestly supersymmetric
way, the interacting (2,0) theory describing the fivebrane near
the boundary of $AdS_7$. In order to do so, we
determine first the realisation of the superconformal algebra 
on vector fields and the invariant metric.

In section 2, we briefly review  the $AdS_7$ metric and
how it arises in eleven-dimensional supergravity.
In section 3, we determine from \cite{GM, par} a realisation of the
superconformal algebra on vector fields. This realisation
does not provide us with the Killing vector fields, as we show in
section 4, 
but will be the zero order approximation to it.    
In section 5, we look for 
a modification of the realisation found in
section 3. We demand that the action of the 
super-Poincar{\'e} group be not modified and we show, in section 6,
that the vector fields have an expansion in the radius, $R$, 
of $AdS$. In this section we also determine explicitly the 
first order expansion which we use in section 7 to
construct the invariant metric. In section 8,
relying on the doubly supersymmetric formalism of the
superembedding approach \cite{howe}, we determine from the 
invariant metric
the equations governing the dynamics of the fivebrane to the 
first
order in $R$; this constitutes a good approximation near the boundary of $AdS$.
Due to our choice of coordinates, we get  manifestly
worldvolume superconformal equations of motions.
At order zero in
$R$ the equations, that is at the boundary of $AdS$,
the equations reduce to the doubleton equations
\cite{ber} and at the next order 
we get interaction terms in the
(2,0) theory.
The equations we get are the supersymmetric 
version of the bosonic ones found for radial coordinates
in \cite{mal, kal} and all bosonic degrees of freedom in
\cite{cla2}.
Our conventions and some of the technical tools
are collected in Appendix A;
the superconformal $OSp(6,2|2)$ algebra is presented in Appendix
B.

\section{AdS and near horizon geometry}
\setcounter{equation}{0}

The $d+1$ dimensional 
Anti-de-Sitter manifold is conveniently described 
as the submanifold of flat $d+2$ dimensional flat manifold
with signature $(-,-,+,\dots+)$ embedded by the equation
\begin{equation}
\phi\chi+\eta_{\mu\nu}X^\mu X^\nu=-R^2,\label{ads}
\end{equation}
where $\eta_{\mu\nu}$ is a $d$-dimensional metric
with signature $(-,+,\dots,+)$ and $R$ is the radius of $AdS_{d+1}$
spacetime. The boundary of $AdS$ is obtained by considering points
at infinity subject to (\ref{ads}). More precisely
define the primed coordinates by $\phi=\lambda\phi',\
\chi'=\lambda\chi,$ and $X^\mu=\lambda X'{}^{\mu}$ and take
the limit $\lambda \rightarrow \infty$ then
the boundary is described by the surface
\begin{equation}
\phi'\chi'+\eta_{\mu\nu}X'{}^\mu X'{}^\nu=0,\label{bads}
\end{equation}
subject to the equivalence relation 
$(\phi',\chi',X'{}^{\mu})\equiv s (\phi',\chi',X'{}^{\mu})$
where $s$ is any real non zero number.
The surface (\ref{bads}) is a compactification of a d-dimensional
Minkowski spacetime: when $\phi'\neq 0$ the coordinates $X'{}^{\mu}$
span $\RR^d$ which
when added to the $\phi'=0$ part provides the compactification,
$S^1\times S^{d-2}$, whose
universal cover is $\RR\times S^{d-1}$.
Convenient  coordinates of $AdS$ are given by
$\phi$ and $x^\mu=RX^\mu/\phi$. These are well defined if 
$\phi\neq 0$ and $x^\mu$ remain finite in the scaling
limit $\lambda \rightarrow\infty$. So the boundary at
$\phi=\infty$ can be parametrised by $x^\mu$. 
In these coordinates the $AdS$ metric takes
the form
\begin{equation}
g=
\frac{\phi^2}{R^2} \eta_{\mu\nu}dx^\mu dx^\nu
+
\frac{R^2}{\phi^2} d\phi d\phi.
\end{equation}
This metric has an apparent singularity at $\phi=0$, however as
noted above, this is merely a coordinate singularity. 
Note that the $\phi=0$ part of $AdS_{d+1}$ is given by
$\RR\times AdS^{E}_{d-1}$ where $AdS^{E}$ is Euclidien
Anti-de-Sitter which is topologically a ball and its boundary
is the sphere $S^{d-2}$. In the following, we shall see
that $\phi$ has a natural interpretation as the radial distance
from a fivebrane source and so the fivebrane is at $\phi=0$.
Note also that by
a change of coordinates $\phi\rightarrow \beta\phi^{\alpha}$ 
the metric
becomes $\beta^2\phi^{2\alpha}/{R^2} dxdx+\alpha^2 R^2 d\phi
d\phi/\phi^2$
which when $\alpha=1/2$ and $\beta^2/R^2=2/R$ will be the form we
shall use in the following.

The eleven-dimensional fivebrane is a solitonic solution to
the eleven-dimensional supergravity low energy equations of motion which preserves 
sixteen supersymmetries. The solution is given by
\begin{equation}
g_{p=5}=H^{-1/3}\eta_{\mu\nu}dx^\mu dx^\nu
+H^{2/3}\eta_{ij}d\phi^i\phi^j,\quad G_{i_1\dots
i_4}=\epsilon_{i_1\dots i_5}\partial^{i_5}H,
\end{equation}
where $x^\mu.\ \mu=0,\dots 5$ are the coordinates along the
fivebrane
and $\phi^i$ are the transverse coordinates. The function
$H$ entering the solution is   radial and harmonic,
$\eta^{ij}\partial_i\partial_j H=0$, and so it reads
\begin{equation}
H= c+
\frac{R^3}{8\phi^3},
\end{equation}
where $\phi^2=\eta_{ij}\phi^i\phi^j$, $c$ is a constant and
\begin{equation}
R^3=Nl_p^3,
\end{equation}
where $l_p$ is the eleven-dimensional Planck length    
and $N$ is the charge of the field strength $G$ or the number of
fivebranes.

If one insists on having an asymptotically flat eleven dimensional
spacetime then one has to impose that $c$ is not zero and one can
set it equal to one. However, another solution is to set $c=0$.
This does not change the charge of the $G$ field strength
and preserves one-half of the eleven-dimensional supersymetries.
In the latter case, $c=0$, the metric takes the form of
$AdS_7\times S^4$. The radial coordinate $\phi$ combines with the
worldvolume coordinates $x^\mu$ to form $AdS_7$.
In fact in this case, the number of supersymmetries 
is greater then
sixteen since the six-dimensional Poincar{\'e} invariance of the
general ($c\neq 0$) case is transformed to a conformal invariance
under the group $SO(6,2)$ which when combined with the sixteen
supersymmetries gives sixteen other special conformal
supersymmetries. 
Another way to get the $AdS$ spacetime is to consider the 
decoupling limit $l_p\rightarrow 0$ and $\phi/l_p^3$
finite \footnote{this guarantees that the tension of the
string obtained from
 the open membrane
stretched between the origin and $\phi$ is finite}
so that
 $\phi \ll R$  and one can neglect the constant $c$ in $H$.
It is this limit that suggested the relation between
the $AdS_{d+1}$ bulk theory and the worldvolume 
$d$-dimensional effective theory \cite{mal}. 

More generally, the $AdS/CFT$ conjecture states that M-theory 
in $AdS_7\times S^4$ is holographically equivalent
to  the six-dimensional $(2,0)$ theory at the boundary of
$AdS$ \cite{wi}.  In the following, we shall show that it is
possible to determine as an expansion in $R^3/\phi$
the metric and isometries of the superbackground whose
bosonic limit is $AdS_7\times S^4$. We shall calculate explicitly
the first order terms which give the correction to the free
dynamics of the fivebrane near the (part of the) boundary at $\phi=\infty$.

\section{A realisation of $Osp(6,2|2)$ on vector fields}
\setcounter{equation}{0}

In this section we determine, from the free 
six-dimensional (2,0) multiplet, 
a realisation of $Osp(6,2|2)$ on eleven-dimensional spacetime.

The six-dimensional (2,0) superspace is spanned by the coordinates
($x^\mu,\theta^\alpha$) where $\mu=0,\dots 5$ and $\alpha$
is a spinorial index in the ($4_+,4$) of $SO(5,1)\times SO(5)$,
$SO(5)$ being the R-symmetry group. 
The ``flat" superconformal transformations 
are defined as in ref. \cite{GM} by the requirement that the flat 
six-dimensional supersymmetric metric 
\begin{equation}
g=\eta_{\mu\nu}E^{\mu}\otimes E^\nu,
\end{equation}
where $E^{\mu}=dx^{\mu}-\bar\theta\Gamma^\mu d\theta$, 
transforms by a scale factor,  that is
\begin{equation}
\delta g=L_{\xi}g=-\alpha g,
\end{equation}
where $\alpha$ is the scale factor.
The realisation of the transformations in superspace was
determined and is given by $\xi=\xi^\mu
E_\mu+\xi^\alpha E_\alpha$. Here, $E_\mu=\partial_\mu$ and
$E_\alpha=D_\alpha=\partial_\alpha-(\bar\theta\Gamma^\mu)_\alpha
\partial_\mu$ are the supersymmetric basis of vectors.
The components of $\xi$ are determined
in terms of $\zeta^\mu$, $\zeta^\alpha$ and $\zeta_{ij}$
which are defined by
\begin{equation}
\zeta_\mu=a_\mu\ +\ a_{[\mu\nu]}x^\nu\ +\ \lambda x_\mu\ +\ 
(x^2\eta_{\mu\nu}-2x_\mu
x_\nu)k^{\nu}
,\label{sro1}
\end{equation}
and
\begin{equation}
	\label{sro2}
\zeta^{\beta}=\epsilon^{\beta}
+ x_\mu
(\Gamma^{\mu}){}^{\beta}{}_{\alpha'}\eta^{\alpha'},
\ \ 
\zeta_{ij}=\frac{\epsilon_{ij}}{4}.
\end{equation}
The parameters $a_\mu,\ a_{\mu\nu},\ \lambda,\ k_\mu,\ 
\epsilon_{ij},\ 
\epsilon^{\hat\alpha}$ and $\eta^{\alpha}$ 
are those of infinitesimal translations, Lorentz transformations,
dilatation, special conformal transformations, rotations,
supersymmetries and fermionic
special conformal transformations.
The six-dimensional super Killing vector fields are given by
\cite{GM}
\begin{eqnarray}
\xi_0^{\mu} & = & 
\zeta^\mu -
2\bar\theta\Gamma^{\mu}
\zeta+\bar\theta
\left(
\Gamma^{\mu i j}\zeta_{ij} 
+{\frac{1}{4}}\,
\Gamma^{\mu\mu_1\mu_2}\partial_{\mu_1}
\zeta_{\mu_2}
\right) \theta
\nonumber\\
&&
+{\frac{1}{2}}
\, \bar\theta\Gamma^{\mu\mu_1\mu_2}\theta\
\bar\theta\Gamma_{\mu_1}\partial_{\mu_2}\zeta
-
{\frac{1}{64}}\,
\bar\theta\Gamma^{\mu\mu_1\mu_2}\theta\
\bar\theta \Gamma_{\mu_1\mu_2\mu_3}\theta\
\partial_{\rho}\partial^{\rho}\zeta^{\mu_3},\label{xim}
\end{eqnarray}
and
\begin{eqnarray}
\xi_0^{\alpha}
& = & 
\left[
\zeta-\Gamma_{ij}\theta\ \zeta^{ij} 
+{\frac{1}{12}}
\theta\ \partial_\mu\zeta^\mu
-{\frac{1}{4}}
\Gamma^{\mu\nu}\theta\ \partial_{\mu}\zeta_{\nu}
\right.\nonumber\\
&&
-{\frac{1}{6}}
\theta\ \bar\theta\Gamma^\mu\partial_\mu\zeta
-{\frac{1}{2}} \Gamma^{\mu\nu}\theta\ 
\bar\theta\Gamma_\mu \partial_\nu\zeta
-{\frac{1}{24}}
\Gamma_{\mu_1\mu_2}\partial_{\mu_3}\zeta\
\bar\theta\Gamma^{\mu_1\mu_2\mu_3}
\theta\nonumber\\
&&\left.
+{\frac{1}{32}}
\Gamma^{\mu_1\mu_2}\theta\
\bar\theta \Gamma_{\mu_1\mu_2\mu_3}\theta\
\partial_{\sigma}\partial^{\sigma}\zeta^{\mu_3}
\right]^{\alpha}
.\label{xia}
\end{eqnarray}
The scale factor, $\alpha$, is given by
\begin{equation}
\alpha={\frac{1}{3}}\partial_\mu\zeta^\mu
-{\frac{2}{3}}\bar\theta\Gamma^\sigma\partial_\sigma\zeta,
\end{equation}
which reads explicitly
\begin{equation}
\alpha=2\lambda-4x^\mu k_\mu-4
\bar\theta\eta.
\end{equation}
The (2,0) on-shell multiplet is described by a superfield
$\Phi^i$ in the vectorial representation of $SO(5)$
subject to the constraint \cite{town1}
\begin{equation}
D_{\alpha }\Phi^{i}={\frac{1}{4}}(\Gamma^{i}_{\ j})
_{ \alpha }^{\  \beta  }D_{\beta }\Phi^{j}.\label{cont}
\end{equation}
It has been shown in \cite{GM} that (\ref{cont})
is invariant under the superconformal group
provided the scalar superfield $\Phi^i$ transforms as
\begin{equation}
	\label{delta0phi}
\delta_0\Phi^i=\xi_0(\Phi^i)+\Lambda^{i}{}_{j}(\xi)\Phi^j,
\end{equation}
where
\begin{equation}
	\label{eq:Lambda}
\Lambda_{ij}(\xi) = 
4\zeta_{ij}-{\frac{2}{3}}
\bar\theta\Gamma_i\Gamma_j\Gamma^\sigma \partial_\sigma\zeta
+{\frac{1}{4}}
\bar\theta\Gamma_\nu\Gamma_{ij}\theta
\partial_\sigma\partial^\sigma\zeta^\nu
+{\frac{1}{3}}\eta_{ij}
\partial_\mu\zeta^\mu.
\end{equation}
From \eqref{delta0phi} and \eqref{eq:Lambda} we get
\begin{equation}
\delta_0 \Phi = \xi_0(\Phi) + \alpha \Phi,
\end{equation}
Where $\Phi^2=\Phi^i\Phi_i$.
The fermionic superfield 
whose $\theta=0$ component, $\psi^{\alpha'}$
is the superpartner of $\phi^i$,
the $\theta=0$ component of $\Phi^i$, is
$\Psi=1/5\Gamma_iD\Phi^i$. Under a superconformal transformation,
it transforms as
\begin{equation}
\delta\Psi=\xi(\Psi)+\alpha\Psi+
\Phi^j\Gamma_j D\alpha+D\xi\Psi,
\end{equation}
where $D\xi$ denotes the matrix $D_\alpha\xi^\beta$.
We can construct a realisation of the superconformal group in
eleven-dimensional superspace as follows.
We consider $x^{\hat \mu}=(x^\mu,\phi^i)$
to be the even coordinates and $\theta^{\hat \alpha}=
(\theta^{\alpha},\psi^{\alpha'})$ to be the odd coordinates.
The two spinors of $SO(5,1)\times SO(5)$ of opposite
six-dimensional chirality combine to form one Majorana spinor
of $SO(10,1)$. 
The vector fields
\begin{equation}
\xi_0=\xi_0^{\mu}\partial_\mu+\xi_0^{\alpha}D_\alpha
-\Lambda^{i}{}_{j}\phi^{j}\partial_i-
\Big((\alpha+\phi^{j}\Gamma_jD\alpha+D\xi)\psi\Big)
^{\alpha'}
\partial_{\alpha'},
\end{equation}
where $\partial_i=\partial/\partial\phi^i$ and
$\partial_{\alpha'}=\partial/\partial \psi^{\alpha'}$,
provide a realisation of the superconformal algebra on eleven
dimensional superspace. Note that $\phi^i$ and $\psi^{\alpha'}$
 are no longer fields but become coordinates in the
eleven-dimensional superspace.

\section{Modified bosonic realisation}
\setcounter{equation}{0}

In this section, we show that the realisation found
in the previous section does not provide us with the super Killing
vectors of $AdS^7\times S^4$. In order to see that, it suffices to
examine the bosonic part of the metric and the transformations.

The bosonic metric of $AdS_7\times S^4$
is given by
\begin{equation}
g=
\frac{2\phi}{R} \eta_{\mu\nu} dx^\mu dx^\nu
+\frac{R^2}{4\phi^2}
\eta_{ij} d\phi^i d\phi^j,
\end{equation}
where $\phi^2=\phi^i\phi_i$.
This metric is not invariant under the transformations generated
by the bosonic part of $\xi_0$ constructed above.
In fact we have
\begin{equation}
\delta_0( \eta_{\mu\nu} dx^\mu dx^\nu)
=-\alpha \eta_{\mu\nu} dx^\mu dx^\nu
\end{equation}
and  
\begin{equation}
\delta_0\phi=\alpha\phi,
\end{equation}
so the first piece of the metric is invariant.
However the second part is not invariant and its variation is
given by
\begin{equation}
\delta_{0}
\left(
\frac{R^2}{4\phi^2}
\eta_{ij} d\phi^i d\phi^j
\right)
=
\frac{R^2}{4\phi^2}\phi^i
\left(
d\alpha d\phi^j+d\phi^j d\alpha
\right)
\eta_{ij}
\end{equation}
This variation can be cancelled by
modifying the transformation of $x^\mu$ by
\begin{equation}
\delta x^{\mu}=\delta_0 x^{\mu}
+
\frac{R^3}{8\phi}
\partial^\mu\alpha.
\end{equation}
The modified transformations, 
as will be seen in the next section, still close on the 
superconformal
algebra.
In order to extend the 
bosonic metric to the superconformal case 
we have to 
 modify  the realisation found in the previous section
 of the superconformal algebra
on eleven dimensional superspace. This will be the goal of the
next section.

\section{Modified superconformal realisations}
\setcounter{equation}{0}

Let $g$ be an element of the superconformal algebra
which is presented in Appendix B,
the vector field which gives a realisation of 
$Osp(6,2|2)$ must verify
\begin{equation}
	\label{eq:closure}
[\xi(g),\xi(g')]=\xi([g,g']),
\end{equation}
In terms of the generators, $\xi(g)$ is of the form
\begin{equation}
\xi=a^\mu P_\mu+\epsilon^\alpha Q_\alpha+ 
\frac{1}{2} a^{\mu\nu}M_{\mu\nu}+
\frac{1}{2} \epsilon^{ij}J_{ij}+k^{\mu}K_{\mu}+
\eta^{\alpha'}S_{\alpha'}+\lambda D,
\end{equation}
We write $\xi$ as $\xi_0+\Delta\xi$,
where $\xi_0$ is the previously determined solution to the
closure relations. So $\Delta\xi$ must be a solution of
\begin{equation}
[\Delta\xi(g),\xi_0(g')]+[\xi_0(g),\Delta \xi(g')]
+[\Delta \xi(g),\Delta\xi(g')]=\Delta\xi([g,g']).
\end{equation}

We require that $(x^\mu, \theta^{\alpha})$ span the 
six-dimensional superspace so that translations and supersymmetric
transformations are unchanged
\begin{equation}
\Delta P_\mu=\Delta Q_{\alpha}=0,
\end{equation}
and we also require that the Lorentz and $R$-group $SO(5)$ 
realisations
be unchanged
\begin{equation}
\Delta M_{\mu\nu}=\Delta J_{ij}=0 .
\end{equation}
The weights of the different coordinates remain the same so
\begin{equation}
\Delta D=0.
\end{equation}
So the modification concerns only $\Delta K_\mu \equiv X_\mu$
and $\Delta S_{\alpha} \equiv X_{\alpha}$. These have to obey the
following constraints which are  the closure relations
\eqref{eq:closure} expressed in terms of the generators
\begin{equation}
\begin{split}
\ [ P_\mu, X_\alpha]  &=  0,\  \ [ P_\mu, X_\nu]\ =\ 0,\\ 
 \ \{ X_\alpha,Q_\beta \} &= 0, \ \ 
[X_\mu,Q_{\alpha}]\ =\ -(\Gamma_{\mu})_{\alpha}{}^{\beta}X_{\beta},
\\ 
\  [ X_\mu, K_\nu] &+ [K_\mu,X_\nu]\ +\ [ X_\mu, X_\nu]\ =\ 0,\\ 
\ [ X_\alpha,S_\beta] &+ [S_\alpha,X_\beta]\ +\ 
\{X_\alpha,X_\beta\}\ =\ 
2(\Gamma^{\mu})_{\alpha\beta}X_{\mu}.
\end{split}
\end{equation}
The other commutation relations give the Lorentz, $SO(5)$
and dilatation
transformations of $X_\mu$ and $X_\alpha$.
The first two equations imply that the components of $X_\mu$ and
$X_\alpha$ are independent of $x$ in the 
($\partial_\mu, \ D_{\alpha},\ \partial_i,\ \partial_{\alpha'})$ basis.
Similarly the next two equations imply that the components of
$X_\alpha$ are independent of $\theta$ and that those of $X_{\mu}$
are of the form
\begin{equation}
X_\mu=Y_\mu+\theta^\alpha
\left(\Gamma_\mu\right)_{\alpha}{}^\beta X_{\beta},
\label{yy}
\end{equation}
where the components of $Y_\mu$ are independent of $\theta$.
The last equation turns out, after the use of Jacobi
identities, to be a consequence of the fifth relation.
Finally we end with the equation 
\begin{equation}
[X_\mu,K_\nu]+[K_\mu,X_\nu]+[X_\mu,X_\nu]=0,
\end{equation}
where the components of $X_{\mu}$ are independent
of $x$ and depend linearly on $\theta$ as in equation (\ref{yy}).
 
Let us examine first the bosonic case, where
$\theta$ and $\psi$ are absent. The most general form
of $X_\mu$ whose components are independent of $x$ and
have the correct transformations under $SO(5,1)$, $SO(5)$
and dilatations is given by
\begin{equation}
X_{\mu}^{(b)}
=
a
\frac{1}{\phi}
\partial_{\mu} ,
\end{equation}
where $a$ is a constant. Note that there are no components on 
$\partial_i$ because there are no Lorentz vectors independent of
$x$. This $X_\mu$ satisfies $[X_\mu,X_\nu]=0$
and a non-trivial calculation gives, for any value of $a$, 
\begin{equation}
[X_\mu^{(b)},K_\nu]+[K_\mu,X_\nu^{(b)}]=0.
\end{equation}
So in the bosonic case we have a one parameter deformation of the
realisation of the conformal Lie algebra on space-time.

In the supersymmetric case, the most general expression for
$X_\mu$ is much more complicated, it has an expansion 
in powers of $\psi$ which a priori stops at sixteen $\psi$'s.
However, this power series in $\psi$ is correlated, as will be
shown in the next section, with
the dependence on $R$ and $\phi$. 
This will help simplifying the resulting
expressions.

\section{Expansion in R}
\setcounter{equation}{0}

Define $\phi'{}^{i}$ by $\phi^i=R^3\phi'{}^{i}$ then the bosonic
metric, expressed in the new coordinates, when
divided by $R^2$ does not depend on $R$.
From the expression of $\xi_0$, we deduce that if we define
$\psi'=R^{-3}\psi$ then $\xi_0$, when expressed in terms of the
primed quantities is independent of $R$. Since the bosonic
metric has only an overall scale dependence 
on $R$ the full $\xi$,
which are the Killing vectors of the full superconformal metric,
are independent of $R$ when expressed in terms of the primed
quantities.

Let us first consider $Y_\mu$,
it has an expansion in $\psi'$ of the form
\begin{multline}
Y_{\mu}
=
\sum_{n}
\left(
\frac{y_{(n)\mu}^\nu (\psi') }{ {\phi'}^{1+{5n/4}} }
\partial_\nu
+
\frac{y_{(n)\mu}^\alpha (\psi') }{ {\phi'}^{3/4+{5n/4}} }
D_\alpha
\right.
\\ 
\left.
+
\frac{y_{(n)\mu}^i(\psi') }{ {\phi'}^{-1/2+{5n/4}} }
\partial'_i
+
\frac{y_{(n)\mu}^{\alpha'}(\psi') }{ {\phi'}^{-3/4+{5n/4}} }
\partial'_{\alpha'}
\right),
\end{multline}
where the $y_{(n)\mu}^{M}(\psi')$ are homogeneous polynomials in
$\psi'$ of degree $n$. 
The powers of $\phi$ in the denominators are determined from the
behaviour under dilatations.
When expressed in terms of 
unprimed quantities we get
\begin{multline}
	\label{Ymu}
Y_{\mu}
=
\sum_{n}
\left(
\r^{4+n}
\frac{y_{(n)\mu }^\nu(\psi) }{ \phi^{1+{5n/4}} }
\partial_\nu
+
\r^{3+n}
\frac{y_{(n)\mu}^\alpha(\psi) }{ \phi^{3/4+{5n/4}} }
D_\alpha 
\right.
\\
\left.
+\r^{2+n}
\frac{y_{(n)\mu }^i(\psi) }{ \phi^{-1/2+{5n/4}} }
\partial_i
+
\r^{1+n}
\frac{y_{(n)\mu }^{\alpha'}(\psi) }{ \phi^{-3/4+{5n/4}} }
\partial_{\alpha'}
\right)
,
\end{multline}
where $\r=R^{3/4}$. Lorentz invariance constrains the 
degrees of polynomials appearing in \eqref{Ymu}.
For example, all terms quadratic in $\psi$ are of the form
$\bar\psi\Gamma^{\mu_1\mu_2\mu_3}\psi,\ 
\bar\psi\Gamma^{\mu i j}\psi,\
\bar\psi\Gamma^{\mu_1\mu_2\mu_3 i}\psi$,
so they all have an odd number of six-dimensional vector indices
and thus cannot contribute to $Y_\mu^\nu$. Using this
kind of arguments one can show that if a polynomial of
degree $n$ contributes to one component then the other polynomials
appearing in the component have degrees $n+4p$.
The lowest degree polynomials appearing in each component are
given by
\begin{equation}
	\label{yymu}
\begin{split}
\ y_{(0)\mu}^{\nu}
&=
 a_1
\frac{\eta_\mu{}^\nu }{ \phi}
\\
\ y_{(1)\mu}^{\alpha}
&= 
a_2 
(\Gamma_{\mu i}\psi)^{\alpha}
\frac{\phi^i}{\phi^3}
+
a_3(\Gamma_{\mu}\psi)^{\alpha}
\frac{1}{\phi^2}
\\
\ 
y_{(2)\mu}^{i}
&= 
a_4\bar\psi\Gamma_\mu{}^{ij}\psi
\frac{\phi_j}{\phi^3}
\\
\ 
y_{(3)\mu}^{\alpha'}
&=
a_5(\Gamma^{\nu_1\nu_2}\psi)^{\alpha'}
(\bar\psi\Gamma_{\mu\nu_1\nu_2}\psi)
\frac{1}{\phi^3}
+
a_6(\Gamma^{\nu_1\nu_2 i}\psi)^{\alpha'}
(\bar\psi\Gamma_{\mu\nu_1\nu_2}\psi)
\frac{\phi_i}{\phi^4}
\\
&+
a_7(\Gamma^{\nu_1\nu_2}\psi)^{\alpha'}
(\bar\psi\Gamma_{\mu\nu_1\nu_2 i}\psi)
\frac{\phi^i}{\phi^4}
+
a_8(\Gamma^{\nu_1\nu_2 i}\psi)^{\alpha'}
(\bar\psi\Gamma_{\mu\nu_1\nu_2 j}\psi)
\frac{\phi_i\phi^j}{\phi^5},
\end{split}
\end{equation}
where $a_i,\ i=1,\dots,8$ are arbitrary constants.
We have used Fierz rearrangements to eliminate some 
of the terms with three powers of $\psi$ (see Appendix A).
Note that the leading power of $r$ in \eqref{Ymu} for all
the components is $\r^4$;
from \eqref{Ymu} and \eqref{yymu} we see that the general form of
$Y_\mu$ is given by
\begin{equation}
Y_\mu=\sum_{p=0}^{4}
\r^{4(p+1)}
Y_{\mu}^{(p)},
\end{equation}
with
\begin{multline}
Y_{\mu}^{(p)}
=
y_{(4p)\mu}^{\nu}
\frac{1}{(\phi)^{1+5p}}
\partial_\nu
+
y_{(1+4p)\mu}^{\alpha}
\frac{1}{(\phi)^{2+5p}}
D_\alpha
\\
+ 
y_{(2+4p)\mu}^{i}
\frac{1}{(\phi)^{2+5p}}
\partial_i
+
y_{(3+4p)\mu}^{\alpha'}
\frac{1}{(\phi)^{3+5p}}
\partial_{\alpha'}.
\end{multline}
Similarly, one can make an analogous analysis for $X_\alpha$.
Here, the lowest degree polynomial contributing to $X_\alpha$ are
given by
\begin{equation}
	\label{xmu}
\begin{split}
x_{(1)\alpha}^{\mu}
&=
b_1(\Gamma^\mu\psi)_{\alpha}
\frac{1}{\phi^2}
+
b_2(\Gamma^{\mu i}\psi)_{\alpha}
\frac{\phi_i}{\phi^3}
\\
x_{(2)\alpha}^{\beta}
&=
b_3 (\Gamma_{\nu ij})_{\alpha}{}^{\beta}
(\bar\psi\Gamma^{\nu ij}\psi)
\frac{1}{\phi^3}
+
b_4 (\Gamma_{\nu ij})_{\alpha}{}^{\beta}
(\bar\psi\Gamma^{\nu ik}\psi)
\frac{\phi^j\phi_k}{\phi^5}
\\
&
+
b_5 (\Gamma_{\nu i})_{\alpha}{}^{\beta}
(\bar\psi\Gamma^{\nu ij}\psi)
\frac{\phi_j}{\phi^4}
+
b_6 (\Gamma_{\nu ijk})_{\alpha}{}^{\beta}
(\bar\psi\Gamma^{\nu ij}\psi)
\frac{\phi^k}{\phi^4}
\\
x_{(3)\alpha}^{i}
&=
b_7(\Gamma_\nu\psi)_{\alpha}
(\bar\psi\Gamma^{\nu i j}\psi)
\frac{\phi_j}{\phi^4}
+
b_8(\Gamma_{\nu j}\psi)_{\alpha}
(\bar\psi\Gamma^{\nu i k}\psi)
\frac{\phi^j\phi_k }{ \phi^5}
\\
x_{(0)\alpha}^{\alpha'}
&=
b_9\delta_{\alpha}{}^{\alpha'}\phi
+b_{10}(\Gamma_i)_{\alpha}{}^{\alpha'}\phi^i.
\end{split}
\end{equation}
The corresponding expansion in $r$ follows
with the difference with respect to $Y_\mu$
that the last term in \eqref{xmu} contributes at order
zero and the contribution to the first order comes from terms
with four $\psi$'s. 
We deduce that $X_\alpha$ and $X_\mu$ have an expansion of the
form
\begin{equation}
X_\alpha=\sum_{p=1}^{4} \r^{4p} X_{\alpha}^{(p)},
\ \ \
X_\mu = \sum_{p=1}^{4} \r^{4p} X_{\mu}^{(p)},
\end{equation}
and the closure relation gives
\begin{equation}
	\label{closure}
[K_\mu,X_{\nu}^{(p)}]+[X_{\mu}^{(p)},K_\nu]
=\sum_{q+q'=p}[X_{\nu}^{(q)},X_{\mu}^{(q')}].
\end{equation}
Since $K_\mu$ has an $x$ and $\theta$ dependence, 
each component of the  equation  above gives upon
identification of terms dependent on $x$ and $\theta$
a series of equations which consistency is not
guaranteed {\it a priori}. To the lowest order, equation 
\eqref{closure}
yields the linear equation
\begin{equation}
[K_\mu,X_{\nu}^{(1)}]+[X_{\mu}^{(1)},K_\nu]=0,
\end{equation}
where $X_\mu^{(1)}=Y_{\mu}^{(1)}+\bar\theta\Gamma_\mu X^{(1)}$,
with the components of $Y_{\mu}^{(1)}$ given in \eqref{yymu}
and those of $X_{\alpha}^{(1)}$ given in \eqref{xmu}.
The detailed analysis of the equations is long but
straightforward. 
The number of equations is much greater than the number of
coefficients. We have checked the compatibility of all the
equations.
Some of the resulting equations are homogeneous 
and are identically
satisfied, if we set $a_1\neq 0$, then the inhomogeneous
equations allow the determination of the other coefficients in
terms of $a_1$. 
We obtain the following results for the
coefficients in $Y_\mu$ :
\begin{equation}
\begin{split}
&
a_3= a_5 = a_6 = a_7 = 0 \\
&
a_1 =- 4 a_2 = 8 a_4 = 32 a_8 
\end{split}
\end{equation}
and for $X_\alpha$
\begin{equation}
\begin{split}
&
b_1 = b_5 = b_6 = b_7 = b_9 = b_{10} = 0 \\
& 
b_2 = -\frac{a_1}{4},\
b_3 = \frac{a_1}{128},\ 
b_4 = -\frac{a_1}{32},\ 
b_8 = -\frac{a_1}{32}.
\end{split}
\end{equation}
%

\section{Invariant metric}
\setcounter{equation}{0}

In this section we shall use the Killing vector fields
determined in the preceding section to deduce the invariant
metric.
We write the metric in the form
\begin{equation}
g=
\frac{2\phi}{R}
\eta_{\mu\nu}\tilde E^{\mu}\otimes \tilde E^\nu
+
\frac{R^2}{4\phi^2}\eta_{ij}\tilde E^i\otimes\tilde E^{j}.
\end{equation}
taking into account the behaviour under translation,
supersymmetry, dilatations, rotation and Lorentz transformations
 as well as
a $\ZZ_2$ symmetry acting
as $\phi^i \to -\phi^i$ and $\psi \to - \psi$,
we get, to order $\r^4$ 
the following expression for 
$\tilde E^\mu$
\begin{eqnarray}
\tilde E^\mu
&=&
E^\mu+\r^4
\left(
c_1
\frac{1}{\phi^5}
(\bar\psi\Gamma^{\mu\nu_1\nu_2}\psi)
(\bar\psi\Gamma_{\nu\nu_1\nu_2}\psi) E^\nu 
+c_2
(\bar\psi\Gamma^\mu d\psi)
\frac{1}{\phi^3}
\right.
\nonumber\\
&&
+c_3
(\bar\psi\Gamma^{\mu ij}\psi)
(\bar\psi\Gamma_{ij}d\theta)
\frac{1}{\phi^4}
+c_4
(\bar\psi\Gamma^{\mu ij}\psi)
(\bar\psi\Gamma_{kj}d\theta)
\frac{\phi_i\phi^k}{\phi^6}
\nonumber\\
&&
+c_5
(\bar\psi\Gamma^{\mu ij}\psi)
(\bar\psi\Gamma_{i}d\theta)
\frac{\phi_j}{\phi^5}
+c_6
(\bar\psi\Gamma^{\mu ij}\psi)
(\bar\psi\Gamma_{ijk}d\theta)
\frac{\phi^k}{\phi^4}
\nonumber\\
&&
+c_7
(\bar\psi\Gamma^{\mu\nu_1\nu_2}\psi)
(\bar\psi\Gamma_{\nu_1\nu_2 i}d\theta)
\frac{\phi^i}{\phi^5}
+c_8
(\bar\psi\Gamma^{\mu\nu_1\nu_2 i}\psi)
(\bar\psi\Gamma_{\nu_1\nu_2}d\theta)
\frac{\phi_i}{\phi^5}
\nonumber\\
&&
\left.
+c_9
(\bar\psi\Gamma^{\nu ij}\psi)
(\bar\psi\Gamma^{\mu}{}_{\nu i}d\theta)
\frac{\phi_j}{\phi^5}
+c_{10}
(\bar\psi\Gamma^{\mu ij}\psi)
\frac{\phi_id\phi_j}{\phi^5}
\right).
	\label{Emu}
\end{eqnarray}
The terms in $c_3,\ c_4,\ c_5$ and $c_6$
are redundant since, by Fierz rearrangements they can be cast
into the form of already present terms (see 
Appendix A); so we can set these coefficients to zero.
Similarly, we get for $\tilde E^i$, to the order zero
\begin{multline}
	\label{Ei}
\tilde E^i
=
d\phi^i
+d_1
(\bar\psi\Gamma^id\theta)
+d_2
(\bar\psi\Gamma^{ij}d\theta)
\frac{\phi_j}{\phi}
\\
+d_3
(\bar\psi\Gamma^{j}d\theta)
\frac{\phi^i\phi_j}{\phi^2}
+d_4
(\bar\psi d\theta)
\frac{\phi^i}{\phi}
+d_5
(\bar\psi\Gamma_{\mu}{}^{ij}\psi)
\frac{\phi_j}{\phi^2}
E^\mu.
\end{multline}
The coefficients $c_i$ and $d_i$ are determined by the requirement
of invariance of the metric under conformal and special conformal
transformations which were determined in the preceding section.
The metric can be decomposed as a term of order zero and a term of
order one
\begin{equation}
g=g_0+g_1,
\end{equation}
where 
\begin{equation}
g_0=
\frac{2\phi}{R}
\eta_{\mu\nu} E^{\mu}\otimes E^\nu
\end{equation}
and 
\begin{equation}
g_1=
\frac{2\phi}{R}
\eta_{\mu\nu}
\Big( 
(\tilde E^{\mu}-E^\mu)\otimes  E^\nu
+
E^\mu\otimes(\tilde E^\nu-E^\nu)
\Big)
+
\frac{R^2}{4\phi^2}
\eta_{ij}\tilde E^i\otimes\tilde E^{j}.
\end{equation}
The equations governing the invariance of the metric give 
at order zero
\begin{equation}
L_{\xi_0}g_0=0,
\end{equation}
which is satisfied by the construction of
$\xi_0$ presented in section 3.
At the next order we get
\begin{equation}
L_{\Delta\xi}g_0+L_{\xi_0}g_1=0.
\end{equation}
This equation is satisfied by construction for all
generators except for conformal and superconformal
transformations (the invariance
under the other transformations are implemented 
in the ansatz \eqref{Emu}--\eqref{Ei}). These  give the values of the constants
appearing in $\tilde E^\mu$:
\begin{equation}
\begin{split}
&
c_3 = c_4 = c_5 = c_6 = c_9 = 0 \\
&
c_1 = -\frac{1}{512},\ 
c_2 = -\frac{1}{32}
\\
&
c_7 = -\frac{1}{16}
+
\frac{\sqrt{3}}{128},
\ c_8 = - \frac{1}{32}
+
\frac{\sqrt{3}}{128}
\\
&
c_{10} = -\frac{3}{64}
+ \frac{\sqrt{3}}{32},
\end{split}
\end{equation}
and those appearing in $\tilde E^i$:
\begin{equation}
\begin{split}
&
d_2=d_3=d_4=0,\\
&
d_1=1,\ 
d_5 = \frac{\sqrt{3}}{4}.
\end{split}
\end{equation}
In addition the value of $a_1$ is fixed as
\begin{equation}
a_1=
\frac{1}{2}.
\end{equation}
%

\section{Interaction in the six-dimensional (2,0) theory}
\setcounter{equation}{0}

In this section we derive the 
equations governing a fivebrane in the $AdS_7\times S^4$
background to the first order in $\r^4$.
This will give the first order 
interaction terms in the (2,0) theory. We shall
use the superembedding approach \cite{howe}
which is particularly convenient for our purposes.
This approach is manifestly worldvolume and target space
supersymmetric. It has been shown that, by fixing the
worldvolume coordinates, it leads to the manifestly 
target space
supersymmetric Green--Schwarz equations for p-branes \cite{howe}.
Here, we shall choose to work in the ``physical" gauge
which leads to  manifestly worldvolume
supersymmetric equations in terms of the superfields $\Phi^i$
and $\Psi$ which appeared in the free (2,0) theory
reviewed in section 3.

In the physical gauge, 
the super worldvolume of the fivebrane is spanned 
by the $(x^\mu, \theta^\alpha)$ coordinates.
The dynamics is described by the superfields
$\Phi^i(x,\theta)$ and $\Psi (x,\theta)$. 
Let $e_\alpha$ be a basis of odd vector fields on
the worldvolume.
And let $e^{\hat \mu}$ be the pull-back of the 
eleven-dimensional even moving basis:
\begin{equation}
e^{\hat \mu}
=
\tilde E^{\hat \mu}_\alpha d\theta^\alpha
+\tilde E^{\hat \mu}_\nu dx^\nu
+ \tilde E^{\hat \mu}_i d\Phi(x,\theta)^i
+\tilde E^{\hat\mu}_{\alpha'}d\Psi^{\alpha'}(x,\theta),
\end{equation}
where $d$ is the six-dimensional exterior derivative
($d=dx^\mu\partial_\mu+d\theta^\alpha\partial_\alpha$).
The equations governing the dynamics of the
fivebrane are simply     
\begin{equation}
	\label{superembedding}
e^{\hat\mu}(e_\alpha)=0.
\end{equation}
These are equivalent to the equations 
obtained from the superembedding approach.
The most general odd vector field on the worlvolume reads
\begin{equation}
e_\alpha
=
\Delta^\beta_\alpha D_\beta
+
\Delta^\mu_\alpha\partial_\mu.
\end{equation}
Thanks to the linearity property of \eqref{superembedding}, it is possible 
to normalise
the $e_\alpha$ such that
$e_{\alpha}(d\theta^\beta)=\delta^\beta_\alpha$,
and $e_\alpha$ can be cast in the form
\begin{equation}
e_\alpha=D_\alpha+\Delta^\mu_\alpha\partial_\mu.
\end{equation}
Note that $\Delta_\alpha^\mu$ is determined from 
\eqref{superembedding} as well as the equation relating $\Phi^i$ to $\Psi$.

The $\mu$ components of the equation \eqref{superembedding}
give
\begin{eqnarray}
0 &=&
\Delta^\mu_\alpha+\r^4
\left(
c_1
\frac{1}{\Phi^5}
(\bar\Psi\Gamma^{\mu\nu_1\nu_2}\Psi)
(\bar\Psi\Gamma_{\nu\nu_1\nu_2}\Psi)
\Delta^\nu_\alpha 
-c_2
(e_{\alpha}(\bar\Psi)\Gamma^\mu \Psi)
\frac{1}{\Phi^3}
\right.
\nonumber\\
&&
+\big[
c_7
(\bar\Psi\Gamma^{\mu\nu_1\nu_2}\Psi)
(\Gamma_{\nu_1\nu_2 i}\Psi)
\frac{\Phi^i}{\Phi^5}
-c_8
(\bar\psi\Gamma^{\mu\nu_1\nu_2 i}\Psi)
(\Gamma_{\nu_1\nu_2}\Psi)
\frac{\Phi_i}{\Phi^5}
\big]_\alpha
\nonumber\\
&&
\left.
+c_{10}
(\bar\Psi\Gamma^{\mu ij}\Psi)
\frac{\Phi_i e_\alpha(\Phi_j)}{\Phi^5}
\right),
	\label{eqmu}
\end{eqnarray}
where we have kept non-zero constants.
The $i$ components of  equation \eqref{superembedding}
give
\begin{equation}
	\label{eqi}
0=e_\alpha(\Phi^i)-d_1(\Gamma^i\Psi)_\alpha
+
d_5
(\bar\Psi\Gamma_{\mu}{}^{ij}\Psi)
\frac{\Phi_j}{\Phi^2}
\Delta^\mu_\alpha.
\end{equation}

Equation \eqref{eqmu} shows that, to the first 
order in $\r^4$, $\Delta^\mu_\alpha$ is given by
\begin{multline}
\Delta^\mu_\alpha
=
-\r^4
\left(
-c_2
(D_{\alpha}(\bar\Psi)\Gamma^\mu \Psi)
\frac{1}{\Phi^3}
+\big[
c_7
(\bar\Psi\Gamma^{\mu\nu_1\nu_2}\Psi)
(\Gamma_{\nu_1\nu_2 i}\Psi)
\frac{\Phi^i}{\Phi^5}
\right.
\\
\left.
-c_8
(\bar\Psi\Gamma^{\mu\nu_1\nu_2 i}\Psi)
(\Gamma_{\nu_1\nu_2}\Psi)
\frac{\Phi_i}{\Phi^5}
\big]_\alpha
+c_{10}
(\bar\Psi\Gamma^{\mu ij}\Psi)
\frac{\Phi_i D_\alpha(\Phi_j)}{\Phi^5}
\right),
\end{multline}
Substituting the expression of $\Delta^\mu_\alpha$
in equation \eqref{eqi}, we get 
\begin{eqnarray}
D_\alpha\Phi^i
&=&
d_1 
(\Gamma^i\Psi)_\alpha+
\r^{4}
\left(-
c_2
(D_{\alpha}(\bar\Psi)\Gamma^\mu \Psi)
\frac{1}{\Phi^3}
+
\big[
c_7
(\bar\Psi\Gamma^{\mu\nu_1\nu_2}\Psi)
(\Gamma_{\nu_1\nu_2 j}\Psi)
\frac{\Phi^j}{\Phi^5}
\right.
\nonumber\\
&&
-c_8
(\bar\Psi\Gamma^{\mu\nu_1\nu_2 k}\Psi)
(\Gamma_{\nu_1\nu_2}\Psi)
\frac{\Phi_k}{\Phi^5}
\big]_\alpha
\nonumber\\
&&
\left.
+c_{10}
(\bar\Psi\Gamma^{\mu kj}\Psi)
\frac{\Phi_k D_\alpha(\Phi_j)}{\Phi^5}
\right)
\left(\partial_\mu\Phi^i
+d_5
(\bar\Psi\Gamma_{\mu}{}^{ij}\Psi)
\frac{\Phi_j}{\Phi^2}
\right).
	\label{Dphi}
\end{eqnarray}
To order zero we have \cite{GM}
\begin{equation}
	\label{Dpsi}
\begin{split}
D_{\beta } \Psi_{\alpha  }
&=
-(\Gamma^{\mu}{}_{i})_{\beta\alpha}
\partial_{\mu}\Phi^{i}
+ {\frac{1}{5}}
 H_{\beta\alpha},\\
 D_{\gamma    }H_{\alpha\beta}
&=
5 (\Gamma^{\mu})_{\gamma\beta} \partial_\mu \Psi_{\alpha  }
+5 (\Gamma^{\mu})_{\gamma\alpha} \partial_\mu \Psi_{\beta  }
\end{split}
\end{equation}
where
\begin{equation}
H_{\alpha\beta}=H_{\mu_1\mu_2\mu_3}
(\Gamma^{\mu_1\mu_2\mu_3})_{\alpha\beta},
\end{equation}
$H_{\mu_1\mu_2\mu_3}$ being a self-dual 3-form.
So to order $\r^4$, we can replace in the right hand side of 
equation \eqref{Dphi} $D\Phi^i$ by
$\Gamma^i\Psi$ and $D\Psi$ by its zero order expression \eqref{Dpsi}.
The equation of motion to the first order becomes
\begin{multline}
D_\alpha\Phi^i
=
(\Gamma^i\Psi)_\alpha+
\r^{4}
\left(
-c_2
(D_{\alpha}(\bar\Psi)\Gamma^\mu \Psi)
\frac{1}{\Phi^3}
+\big[
c'_7
(\bar\Psi\Gamma^{\mu\nu_1\nu_2}\Psi)
(\Gamma_{\nu_1\nu_2 j}\Psi)
\frac{\Phi^j}{\Phi^5}
\right.
\\
\left.
-c'_8
(\bar\Psi\Gamma^{\mu\nu_1\nu_2 k}\Psi)
(\Gamma_{\nu_1\nu_2}\Psi)
\frac{\Phi_k}{\Phi^5}
\big]_\alpha
\left(\partial_\mu\Phi^i
+d_5
\bar\Psi\Gamma_{\mu}{}^{ij}\Psi
\frac{\Phi_j}{\Phi^2}
\right)
\right),
\end{multline}
where $c'_7 = c_7 -c_{10}/4$ and $c'_8 = c_8 -c_{10}/4$.
These equations are invariant under the superconformal
transformations which read 
\begin{equation}
\delta\Phi^i
=
\delta_0\Phi^i+a_1 \r^4
\left(
\frac{k^\mu}{\Phi}
-\frac{1}{4\Phi^3}
(\bar\eta'\Gamma^{\mu j}\Psi)
\Phi_j
\right)
\left(
\partial_\mu\Phi^i
-\frac{1}{8} 
(\bar\Psi\Gamma_{\mu}{}^{ij}\Psi)
\frac{\Phi_j}{\Phi^2}
\right)
\end{equation}
where $\eta'=\eta-k_\mu\Gamma^\mu\theta$ and $\delta_0\Phi^i$
is given in \eqref{delta0phi}.

On the boundary of $AdS$, the resulting equations are the
doubleton equation of the free multiplet, the first 
correction in $R$ yields an  interacting (2,0) theory
with a non-linear realisation of the superconformal algebra.

\appendix

\section{Conventions and Fierz rearrangements}
\setcounter{equation}{0}

The eleven dimensional superalgebra reads
\begin{equation}
\{Q_{\hat \alpha},Q_{\hat \beta}\}= 2(\Gamma^{\hat \mu}C)_{\hat
\alpha \hat\beta}P_{\hat\mu},\label{onze}
\end{equation}
where $\hat\alpha=1,\dots, 32,\ \hat\mu=0,\dots 10$ and
$C_{\hat\alpha\hat\beta}$ is an 
antisymmetric matrix verifying
\begin{equation}
C^{-1}\Gamma^{\hat\mu}C=-\Gamma^{\hat\mu T}.
\end{equation}
The reality condition on 11D fermions reads
\begin{equation}
\Psi=C\bar\Psi^{T},
\end{equation}
or equivalently
\begin{equation}
\bar\Psi^{\hat\alpha}=C^{\hat\alpha\hat\beta}\Psi_{\hat\beta}
\equiv \Psi^{\hat\alpha},
\end{equation}
where $C^{\hat\alpha\hat\beta}$ is the inverse of 
$C_{\hat\alpha\hat\beta}$.  We shall use $C$ to raise
and lower indices and the notation
$(\Gamma^{\hat\mu})_{\hat\alpha\hat\beta}$ for  
$(\Gamma^{\hat\mu} C)_{\hat\alpha\hat\beta}$. 

We are interested in the $6+5$ splitting the eleven dimensional spacetime.
A representation of the Gamma matrices is conveniently given by
\begin{equation}
\Gamma^{\mu}=\gamma^{\mu}\otimes 1, \quad
\Gamma^{5+i}={\tilde \gamma} \otimes \gamma^{i},
\end{equation}
where $\mu$ and $i$ are respectively  six and five 
dimensional vector indices, and ${\tilde \gamma}$ is the chirality matrix
in six dimensions:
\begin{equation}
{\tilde \gamma}= \gamma^{0}\dots\gamma^{5}.
\end{equation}
This matrix allows to decompose 11D Majorana fermions
into two 6D $Sp(2)$ symplectic Majorana Weyl fermions:
$\psi^{\hat \alpha} = (\psi^\alpha, \psi^{\alpha'})$. 
We shall also denote $\Gamma^{5+i}$ by $\Gamma^i$.
In this representation the  charge conjugaison matrix $C$ 
may be written as
\begin{equation}
C={\cal C}\otimes\Omega,
\end{equation}
where $\cal C$ is symmetric and verifies
\begin{equation}
{\cal C}^{-1}\gamma^{\mu}{\cal C}=-\gamma^{\mu\ T},
\end{equation}
whereas $\Omega$ is antisymmetric and verifies
\begin{equation}
\Omega^{-1}\gamma^i\Omega=\gamma^{i\ T}.
\end{equation}

The antisymmetrised product of $n$ gamma
matrices is denoted  by
$\Gamma^{\hat\mu_1\dots\hat\mu_n}$. We have
\begin{equation}
(\Gamma^{\hat\mu_1\dots\hat\mu_n}C)^{T}=-(-1)^{n(n+1)/2}
\Gamma^{\hat\mu_1\dots\hat\mu_n}C
.
\end{equation}
A useful relation is the Fierz rearrangement formula which
reads for four Weyl-Majorana fermions \footnote{ 
The summation over the indices $\mu$ and $i$ is ordered in 
the formula \eqref{formuleFierz}~: 
$\mu_1 < \ldots < \mu_{n_1}$
and $i_1 < \ldots < i_{n_2}$. Otherwise some factorials 
appear in the normalization.
}
\begin{eqnarray}
	\label{formuleFierz}
&&\left( \bar\epsilon_1\Pi^{+}\epsilon_2 \right)
\left( \bar\epsilon_3\Pi^{+}\epsilon_4 \right)
=
\nonumber \\
&&
\hskip1cm
-\sum_{\substack{n_1=0,2\\n_2=0,1, 2}} 
{\frac{2}{c_{n_1}\tilde c_{n_2}}}
\left( \bar\epsilon_1
\Gamma^{\mu_1\dots\mu_{n_1}}
\Gamma^{i_1\dots i_{n_2}}\Pi^+\epsilon_4 \right)
\left( \bar\epsilon_3
\Gamma_{\mu_1\dots\mu_{n_1}}
\Gamma_{i_1\dots i_{n_2}}\Pi^+\epsilon_2 \right)
,\nonumber\\
&&
\left( \bar\epsilon_1 \Pi^{+}\epsilon_2 \right)
\left( \bar\epsilon_3\Pi^{-}\epsilon_4 \right)
 = 
\nonumber\\
&&
\hskip1cm
\sum_{n_2=0,1,2} 
{\frac{(-1)^{n_2}}{8\tilde c_{n_2}}}
\left(
-2 \left( \bar\epsilon_1\Gamma^{\mu_1}\Gamma^{i_1\dots i_{n_2}}
\Pi^- \epsilon_4 \right)
\left( \bar\epsilon_3\Gamma_{\mu_1}\Gamma_{i_1\dots i_{n_2}}
\Pi^+ \epsilon_2 \right) \right.
\nonumber\\
&&
\hskip1cm
+\left. \left( \bar\epsilon_1\Gamma^{\mu_1\mu_2\mu_3}
\Gamma^{i_1\dots i_{n_2}}\Pi^-
\epsilon_4 \right)
\left( \bar\epsilon_3\Gamma_{\mu_1\mu_2\mu_3}
\Gamma_{i_1\dots i_{n_2}}
\Pi^+\epsilon_2
\right)
\right)
\end{eqnarray}
with coefficients $c_{n}$ and $\tilde c_{n}$ given by
\begin{eqnarray}
c_{n} = 8\, (-1)^{n(n-1)/2} \ \mbox{ and } \
\tilde c_{n} =  4 \, (-1)^{n(n-1)/2}.
\end{eqnarray}
$\Pi^\pm$ are the chirality projection operators in six 
dmensions.
The formula \eqref{formuleFierz} has been used in our 
derivations to restrict the number of 
independent terms appearing in the 
Killing vectors and in the metric, 
particularly those involving three fermions $\psi$.
First, in the expression \eqref{yymu} of $y^{\alpha'}_{(3)\mu}$ 
and in the expression 
\eqref{Emu} of ${\tilde E}^\mu$, Fierz rearrangements eliminate 
terms with combinations like 
$(\Gamma^i \psi)({\bar \psi} \Gamma_{\nu ji}\psi)$
or $(\Gamma^{j_1 j_2}\psi)({\bar \psi} 
\Gamma_{\nu j_1j_2}\psi)$, indeed
\begin{equation}
\begin{split}
&
(\Gamma^j \psi)({\bar \psi} \Gamma_{\nu ji}\psi)
= 
\frac{1}{4} 
(\Gamma_{i\sigma_1\sigma_2}\psi) 
({\bar \psi} \Gamma_{\nu}{}^{\sigma_1\sigma_2}\psi)
+
\frac{1}{4} 
(\Gamma^{\sigma_1\sigma_2}\psi) 
({\bar \psi} \Gamma_{i\nu\sigma_1\sigma_2}\psi)
\\
&
(\Gamma^{j_1 j_2}\psi)({\bar \psi} \Gamma_{\nu j_1j_2}\psi)
=
-
(\Gamma^{\sigma_1\sigma_2}\psi) 
({\bar \psi} \Gamma_{\nu\sigma_1\sigma_2}\psi)
.
\end{split}
\end{equation}
The expression \eqref{xmu} of $x^i_{(3)\alpha}$ involves 
terms of the form
\begin{equation}
	\label{fierz3psi}
(\Gamma_\sigma N^i_{[i_1 i_2]} \psi) ({\bar \psi} 
\Gamma^{\sigma i_1 i_2} \psi)
.
\end{equation}
There are {\it a priori} twelve terms that can be 
constructed using $\phi^i$ and gamma matrices~:
\begin{equation}
\begin{split}
N^i_{[i_1 i_2]}
= 
& 
a(\delta^i_{i_1} \phi_{i_2} - \delta^i_{i_2} \phi_{i_1}) 
\mathbf{1}
+
a'(\delta^i_{i_1} \phi_{i_2} - \delta^i_{i_2} \phi_{i_1}) 
\frac{\phi_m}{\phi} \Gamma^m 
\\
&
+
b \phi (\delta^i_{i_1} \Gamma_{i_2} - \delta^i_{i_2} 
\Gamma_{i_1})
+
b' (\delta^i_{i_1} \Gamma_{i_2m} - \delta^i_{i_2} \Gamma_{i_1m}) 
\phi^m
\\
&
+
c \frac{\phi^i}{\phi} (\Gamma_{i_1} \phi_{i_2} - \Gamma_{i_2} 
\phi_{i_1})
+
c' \frac{\phi^i \phi^m}{\phi^2} (\Gamma_{mi_1} \phi_{i_2} - 
\Gamma_{m i_2} \phi_{i_1})
\\
&
+
d \phi^i \Gamma_{i_1 i_2}
+
d' \frac{\phi^i \phi^m}{\phi} \Gamma_{m i_1 i_2}
\\
&
+
e (\Gamma^i{}_{i_1}\phi_{i_2} - \Gamma^i{}_{i_2}\phi_{i_1})
+
e' (\Gamma^i{}_{m i_1}\phi_{i_2} - \Gamma^i{}_{m i_2}\phi_{i_1}) 
\frac{\phi^m}{\phi}
\\
&
+
f \phi \Gamma^i{}_{i_1 i_2}
+ f' \Gamma^i{}_{i_1 i_2 m} \phi^m
.
\end{split}
\end{equation}
However a Fierz rearrangement insures that
$N^i_{[i_1 i_2]}$ and $(Q N)^i_{[i_1 i_2]}$ 
where $Q $ is defined as
\begin{equation}
(Q N)^i_{i_1 i_2} =
-\frac{1}{6} N^i_{[j_1 j_2]} \Gamma^{j_1 j_2}{}_{i_1 i_2}
-\frac{1}{3}
\left(
N^i_{[i_1 j]} \Gamma^j{}_{i_2} - N^i_{[i_2 j]} \Gamma^j{}_{i_1}
\right)
,
\end{equation}
are equivalent in the sense that
the expression \eqref{fierz3psi}
computed with $N^i_{[i_1 i_2]}$ and
$(Q N)^i_{[i_1 i_2]}$ are equal.
Examining  the eigenvalues of the operator $Q$ allows to conclude
that
\begin{equation}
\begin{split}
&
b \sim c \sim c' \sim d \sim d' \sim f \sim 0 \\
&
a \sim b' \sim -e \sim - f' \\
&
a' \sim e'
.
\end{split}
\end{equation}
Finally, we end up with only the two terms written in \eqref{xmu}.

\section{The $Osp(6,2|2)$ algebra} 
\setcounter{equation}{0}

\begin{align}
&[ P_\mu , P_\nu ] = 0 \\
&[ P_\mu , M_{\rho\sigma} ] = 
-\eta_{\mu\rho} P_\sigma + \eta_{\mu\sigma} P_\rho \\
&[ P_\mu , J_{ij} ] = 0 \\
&[ P_\mu , D ] =  P_\mu \\
&[ P_\mu , K_\nu ] = -2 \eta_{\mu\nu} D - 2 M_{\mu\nu} \\
&[ P_\mu , Q_{\alpha^\prime} ] = 0\\
&[ P_\mu , S_{\alpha} ] = - {{(\Gamma_\mu)}_{\alpha}}^{\beta^\prime} Q_{\beta^\prime} \\
&[ M_{\mu\nu} , M_{\rho\sigma} ] = 
-\eta_{\mu\sigma} M_{\nu\rho}
+\eta_{\mu\rho} M_{\nu\sigma}
-\eta_{\nu\rho} M_{\mu\sigma}
+\eta_{\nu\sigma} M_{\mu\rho}\\
&[ M_{\mu\nu} , J_{ij} ] = 0 \\
&[ M_{\mu\nu} , D ] = 0 \\
&[ M_{\mu\nu} , K_\sigma ] = \eta_{\mu\sigma} K_\nu - \eta_{\nu\sigma} K_\mu \\
&[ M_{\mu\nu} , Q_{\alpha^\prime} ] = \frac{1}{2} 
{{(\Gamma_{\mu\nu})}_{\alpha^\prime}}^{\beta^\prime} Q_{\beta^\prime} \\
&[ M_{\mu\nu} , S_{\alpha} ] = \frac{1}{2} 
{{(\Gamma_{\mu\nu})}_{\alpha}}^{\beta} S_{\beta} \\
&[ J_{ij} , J_{kl} ] =
\eta_{il} J_{jk} - \eta_{ik} J_{jl} + \eta_{jk} J_{il} - \eta_{jl} J_{ik} \\
&[ J_{ij} , D ] = 0 \\
&[ J_{ij} , K_\mu ] = 0 \\
&[ J_{ij} , Q_{\alpha^\prime} ] = -\frac{1}{2}
{{(\Gamma_{ij})}_{\alpha^\prime}}^{\beta^\prime} Q_{\beta^\prime} \\
&[ J_{ij} , S_{\alpha} ] = -\frac{1}{2}
{{(\Gamma_{ij})}_{\alpha}}^{\beta} S_{\beta} \\
&[ D , D ] = 0 \\
&[ D , K_\mu ] = K_\mu \\
&[ D , Q_{\alpha^\prime} ] = -\frac{1}{2} Q_{\alpha^\prime} \\
&[ D , S_{\alpha} ] = \frac{1}{2} S_{\alpha} \\
&[ K_\mu , K_\nu ] = 0 \\
&[ K_\mu , Q_{\alpha^\prime} ] = - {{(\Gamma_\mu)}_{\alpha^\prime}}^{\beta} S_{\beta} \\
&[ K_\mu ,  S_{\alpha} ] = 0 \\
&{\{} Q_{\alpha^\prime} , Q_{\beta^\prime} {\}} = 2  
{{(\Gamma^\mu)}_{\alpha^\prime \beta^\prime}} P_\mu \\
&{\{} Q_{\alpha^\prime} , S_{\beta} {\}} = 
2 {\cal C}_{\alpha^\prime \beta} + 2 {{(\Gamma^{ij})}_{\alpha^\prime \beta}} J_{ij}
+ {{(\Gamma^{\mu\nu})}_{\alpha^\prime \beta}} M_{\mu\nu} \\
&{\{} S_{\alpha} , S_{\beta} {\}} =
2 {{(\Gamma^\mu)}_{\alpha \beta}} K_\mu 
\end{align}


\end{document}